# Evaluating If Trust and Personal Information Privacy Concerns Are Barriers to Using Health Insurance That Explicitly Utilizes AI

Alex Zarifis , Peter Kawalek & Aida Azadegan

Routledge
Taylor & Francis Group

OPEN ACCESS

# Evaluating If Trust and Personal Information Privacy Concerns Are Barriers to Using Health Insurance That Explicitly Utilizes AI

Alex Zarifis[a], Peter Kawalek[a], and Aida Azadegan[b]

[a]School of Business and Economics, Loughborough University, Loughborough, UK; [b]Intelligent Systems Research Laboratory, University of Reading, Reading, UK

**ABSTRACT**

Trust and privacy have emerged as significant concerns in online transactions. Sharing information on health is especially sensitive but it is necessary for purchasing and utilizing health insurance. Evidence shows that consumers are increasingly comfortable with technology in place of humans, but the expanding use of AI potentially changes this. This research explores whether trust and privacy concern are barriers to the adoption of AI in health insurance. Two scenarios are compared: The first scenario has limited AI that is not in the interface and its presence is not explicitly revealed to the consumer. In the second scenario there is an AI interface and AI evaluation, and this is explicitly revealed to the consumer. The two scenarios were modeled and compared using SEM PLS-MGA. The findings show that trust is significantly lower in the second scenario where AI is visible. Privacy concerns are higher with AI but the difference is not statistically significant within the model.

**KEYWORDS**
Artificial Intelligence; health; information privacy; insurance; trust

## Introduction

Artificial intelligence (AI), particularly machine learning and deep learning is disrupting insurance and health. In insurance natural language processing is utilized extensively by virtual agents interacting with the consumer and AI is also used to detect fraudulent claims (Wang and Xu 2018). In health it is used to make a diagnosis and treatment (He et al. 2019). The use of AI creates value by offering insight and turning insight into action. This happens for several services across several channels and across the whole value chain. Humans still have an important role, but they need support to be able to utilize vast amounts of data and respond quickly. AI is broad and immature in relation to its potential, so this is a challenging area. This is an interdisciplinary topic that goes beyond the narrow focus of developing the necessary technology and testing the usability. Its interdisciplinary nature is comparable to the emergence of other technologies, such as blockchain. There are many questions that need answers, but these questions are heavily influenced by the present capability and the extant diffusion of the technology. The benefits of AI for the consumer in health insurance include customized offers and real-time adaptation of fees. Currently the interface between the consumer purchasing health insurance and AI raises some barriers such as trust and Personal Information Privacy Concern (PIPC) (Gulati, Sousa, and Lamas 2019). The consumer is not a passive recipient of the increasing role of AI. Many consumers have beliefs on what this technology should do. Furthermore, regulation is moving toward making it necessary for the use of AI to be explicitly revealed to the consumer (European Commission 2019). Therefore, the consumer is an important stakeholder and their perspective should be understood and incorporated into future AI solutions in health insurance. This research identified two scenarios, one with limited AI that is not in the interface, whose presence is not explicitly revealed to the consumer and a second scenario where there is an AI interface and AI evaluation, and this is explicitly revealed to the consumer.

The insights AI offers can be summarized into optimization, search and recommendation, and diagnosis and prediction. In addition to the improving technology of AI, the capabilities are also increasing because of the technologies it interacts with. These technologies include big data, internet of things, increased computing capabilities and blockchain (Riikkinen et al. 2018). In general, there is far more data and more capabilities to analyze it. This raises the question whether the impact of AI can be evaluated and guided in isolation or if all these technologies should be evaluated as a new context. Blockchain technology can support the internet of things in terms of security and integrity of the data, the internet of things creates far more data, big data and AI need to analyze it and they need more computing capabilities.

The challenges to AI depend on the specific implementation, the information system it is part of and the specific context. One challenge is implementations of AI that have negative impacts, for example on individuals' health (He et al. 2019). The risks caused by AI seem to come from either using its capabilities to do something harmful more effectively or by AI making incorrect evaluations. In a fully automated system, mistakes will be implemented directly. If AI is working with humans, the humans may act based on the incorrect evaluation. As AI can be unpredictable this raises some questions in terms of control and how to manage the risks. Society, governments and other institutions like the European Union are attempting to regulate and offer guidelines on how to move forward with AI in an ethical way reducing the risks to the u consumer (European Commission 2019). This research focuses on the barriers from the consumer's perspective when they are purchasing health insurance online.

AI offers some unique capabilities but most of the impact is as a part of a wave of innovation that will optimize and create new products, services, business models and business ecosystems. AI can be seen as the catalyst because it can harness the breadth of hardware and software in a way that was not possible before: It can mask the complexity and provide the value to the health insurance consumer. This increased role of AI, and the ecosystem of technologies it utilizes, influence the consumer's attitude. The consumer will interpret some capabilities AI offers as enablers and some as risks and concerns. For example, limited trust and PIPC may be barriers.

The change from related technologies and other trends in society like greener living mean people want to see different principles and values from their insurers. Therefore, the new ethical, privacy and trust challenges AI brings can be approached as part of a holistic reevaluation of the relationship between a consumer and their health insurer. New business models may require a new ethical perspective. Ethics and regulation are evolving as the uses and business models of AI evolve. The new way of interacting with the consumer, the new interfaces or even business models must consider the enablers and barriers to AI in health insurance from the consumer's perspective.

The next section will review health insurance, trust and PIPC to provide the theoretic foundation. This will be followed by the methodology that explains how the model is tested in the two scenarios, with and without visible AI. Finally, the analysis and the conclusion are presented.

## Theoretical background

AI in health insurance raises many new issues grounded in the existing, and widely validated, constructs of ease of use, usefulness, trust and Personal Information Privacy Concern (PIPC).

### Perceived ease of use and usefulness in health insurance with AI

Information systems have been divided into hedonic and utilitarian (van der Heijden 2004). Purchasing health insurance is mostly utilitarian as it is something useful but not necessarily an enjoyable process and the motivation to do it is not enjoyment. Adoption and attitude toward the use of a technology is influenced by some factors that are similar across different contexts and some that are different depending on the context.

Perceived ease of use and perceived usefulness (Davis 1989; Venkatesh, Thong, and Xu 2016) have been found to influence adoption and use across several contexts such as adopting insurance (Lee, Tsao, and Chang 2015). In systems that are more focused on hedonism, a construct for enjoyment could be included along with ease of use and usefulness, but here it is not included. While these measures have been validated in the insurance sector the items used to operationalize them need to be adapted to capture the increased role of AI in health insurance. AI in health insurance can influence perceived ease of use with a personal assistant in several ways. Personal assistants are used in the interaction with the consumer and utilize machine learning for natural language processing and analyzing the consumer request. Firstly, the system using AI and a personal assistant can keep a constant state across many consumer queries. This means the system will keep all the relevant information together and utilize it for each query, so the consumer needs to make less effort. A second example is that AI can interpret unstructured data like pictures and emails from the consumer, so that information does not need to be reentered manually. AI in health insurance can influence perceived usefulness positively by processing applications fast and making customized offers. Some insurers offer customized quotes in minutes (Baloise 2019).

### Trust in health insurance with AI

Trust is important where there is an exchange of value that involves some risk. The higher the risk the more important trust becomes. Making a purchase online is perceived to have higher risk compared to face to face and therefore the role of trust is more significant (McKnight and Chervany 2001). There is an additional risk in purchasing health insurance online because the consumer is taking a risk that the insurer will fulfill their duty and cover the consumer if they make a claim. Insurance companies and the information systems they use have developed sufficiently to build trust in the typical current scenario with limited AI that is not visible to the consumer. The increasing role of AI across the health insurance supply-chain brings new sources of distrust that come from the lack of human attributes in more stages of the supply chain, both front-end and back-end and the real or perceived unpredictability of AI. The consumer's trust in a virtual agent they are interacting with can be reduced by the lack of human factors like visual and auditory emotions (Torre, Goslin, and White 2020). As Figure 1 illustrates the interaction of the consumer can be divided into three scenarios: Firstly, a traditional face to face interaction without utilizing technology. Secondly, a digital interaction online that still utilizes human logic primarily and uses AI in a limited way that is not visible to the consumer. Lastly, the third scenario involves the consumer interacting purely with an AI interface and with the underlying logic and decision making also based on AI.

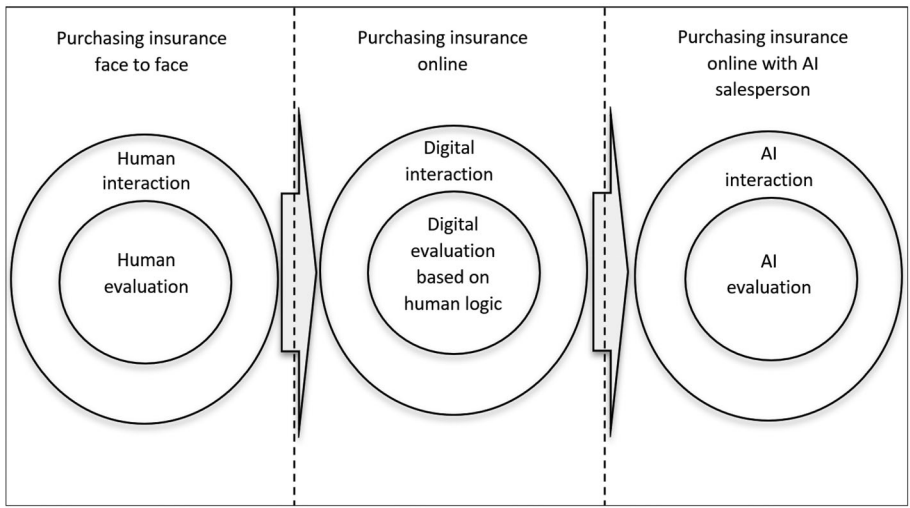

Figure 1. Trust face to face, online, online with AI.

The consumer's trust online is influenced by the psychology of the consumer and sociological influences. The psychology of the individual consumer may give them a propensity to trust (Kim and Prabhakar 2004) which is similar to a disposition to trust (Zarifis et al. 2014) and a trusting stance (McKnight et al. 2011). The social dimension of trust includes institution based trust that covers structural assurance and situational normality (McKnight and Chervany 2001). Structural assurance refers to guarantees, seals of approval and protection from the bank or card used to make the transaction (Sha 2009). Situational normality can include reviews and conforming to social norms.

The risk the consumer perceives from AI having a significant and decisive role in every aspect of their health insurance can lower trust. There is limited familiarity with this level of AI, the ethics of AI are unclear, AI can be unpredictable, the transparency can be limited and the control the humans in the insurance company have over AI is unclear.

## Personal information privacy concerns (PIPC) from threats by AI in health insurance

When a consumer enters personal information such as their date of birth, their bank account details and medical information on their health to acquire health insurance online there is a concern in how this information is used, shared and stored securely. For a consumer to purchase health insurance they need to provide an extensive amount of information;

enough for a criminal to commit fraud. Even if misuse or a security breach is revealed to the consumer, they cannot change their health information like they can change their bank account details to protect their privacy. Therefore, sharing this health information causes privacy concerns (Bansal, Zahedi, and Gefen 2010). These privacy concerns can be a key predictor of using health services online (Park and Shin 2020). Privacy concerns can be divided into perceived privacy control and perceived privacy risks (Dinev et al. 2013). Perceived privacy control can include confidentiality, secrecy and anonymity. Perceived privacy risks can include the sensitivity of the information and the level of regulation (Dinev et al. 2013).

Using AI in healthcare insurance can reduce the perceived information control and increase the perceived risk. The lack of understanding of the role of AI in this context, the unpredictability of AI, the low transparency on the algorithms, the lack of humanness and the unclear ethics may increase the concerns over personal information privacy. Furthermore, AI is part of an ecosystem of technologies such as big data that enhance each other's capabilities and pervasiveness. These increased capabilities and pervasiveness can elevate privacy concerns. When a consumer enters their personal information during the process of acquiring health insurance, they may be thinking about how this personal information could be used against them in the future. Currently, insurance companies offer privacy assurances and privacy seals that can reduce privacy concern (Hui, Teo, and Sang-Yong 2007) however these do not explicitly cover the role of AI.

## Research model and hypotheses

The online consumer evaluates the advantages and disadvantages of purchasing a product or service online before moving forward. Some weaknesses, or even threats, can be overlooked if the advantages are enough. An example of this is the privacy paradox (Norberg, Horne, and Horne 2007) where consumers still give their personal information despite their concerns. Therefore, it is necessary to identify all the relevant factors and model their relationship. The literature review identified two enablers and two barriers to the use of AI in insurance from the consumer's perspective. The previous section on trust also identified how trust is influenced by the humanness of the interface with the insurer, and the humanness of the logic that leads to the evaluation of the insurance offered. As the consumer will be explicitly informed about their interaction with AI (European Commission 2019) this may influence their attitude and raise the barriers of insufficient trust and information privacy concerns. To better understand the scenario where the consumer purchases health insurance with an AI interface and AI logic and decision-making, this new scenario needs to be contrasted with the typical existing scenario. In the existing scenario there is limited AI, that is not in the interface, and its presence is not explicitly revealed to the consumer. Therefore, as illustrated in Figure 2, two scenarios are contrasted, $ß_1$, with limited AI, not explicitly revealed to the consumer and $ß_2$ with an AI interface and AI evaluation explicitly revealed to the consumer.

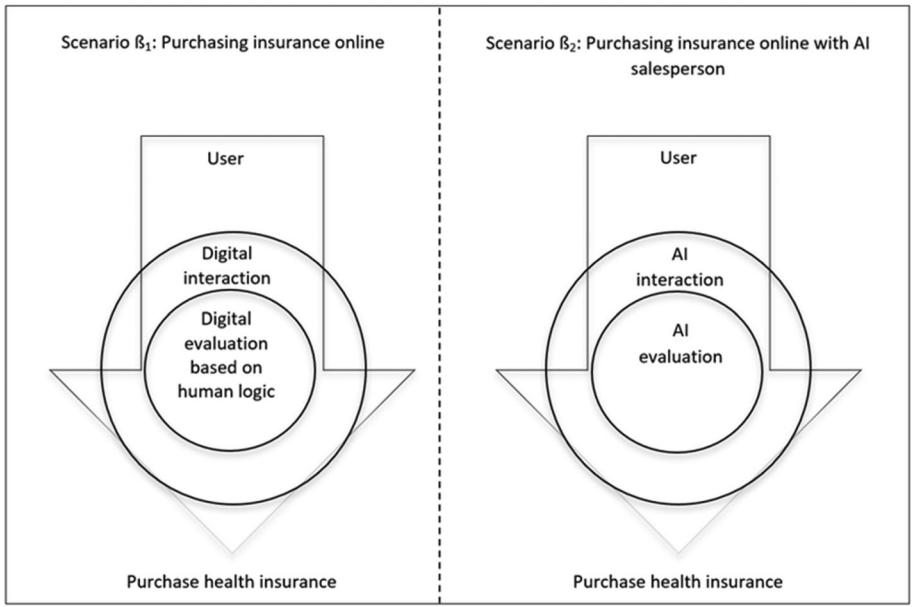

Figure 2. Research approach.

For these scenarios to be tested they are modeled as illustrated in Figure 3. The seven hypotheses proposed cover the relationships between the five constructs. The first two constructs are the enablers, firstly the additional ease of use offered by AI and secondly the additional usefulness offered by AI. The second two constructs are the two barriers of trust and privacy concern. The final construct is the decision to purchase health insurance online.

The literature suggests that additional layers of technology, additional capabilities of technology and a reduction in the role of humans in a process can be perceived as an increase in the risk and therefore a reduction in trust and an increase in information privacy concern (McKnight et al. 2011). Therefore, the first and second hypotheses are the following:

> H1: The consumer will have lower trust if the use of AI is visible to them during the process of purchasing health insurance online.
>
> H2: The consumer will have higher Perceived Information Privacy Concern if the use of AI is visible to them during the process of purchasing health insurance online.

The literature suggests that perceived ease of use has a positive influence on perceived usefulness, trust, privacy concern and the purchase of health insurance in both scenarios (Gefen, Karahanna, and Straub 2003; Al-Khalaf and Choe 2020). Similarly, perceived usefulness has a positive influence on the purchase in both scenarios. Therefore, the final five hypotheses are the following:

> H3: Perceived ease of use will have the same influence on trust in AI if AI is visible during the purchase of health insurance online.

> H4: Perceived ease of use will have the same influence on Personal Information Privacy Concern from AI if AI is visible during the purchase of health insurance online.

> H5: Perceived ease of use will have the same influence on perceived usefulness of AI if AI is visible during the purchase of health insurance online.

> H6: Perceived usefulness will have the same influence on the purchase of health insurance online if AI is visible.

> H7: Perceived ease of use will have the same influence on the purchase of health insurance online if AI is visible.

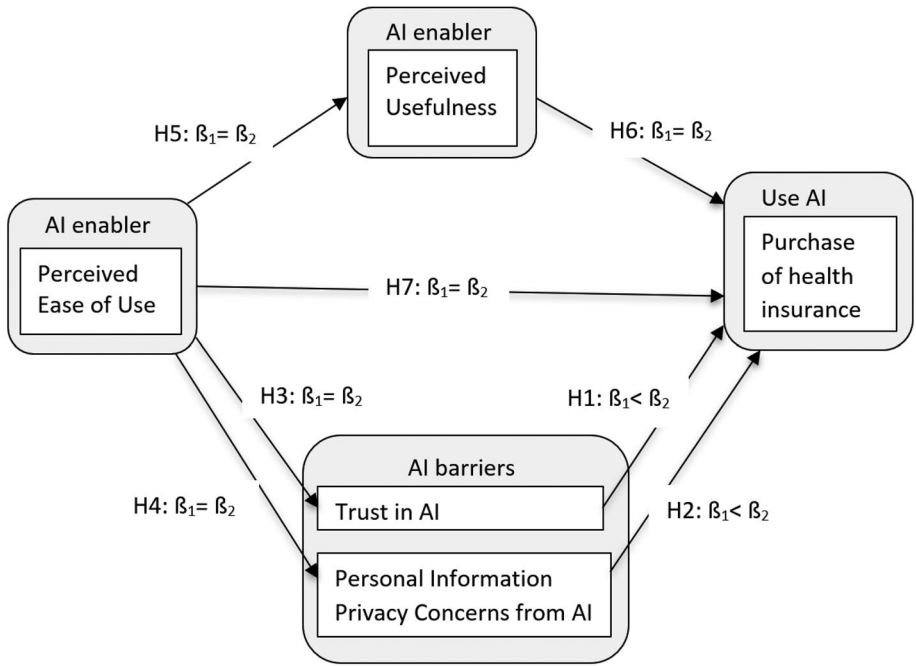

Figure 3. Research model and hypotheses.

## Research methodology

### Measures

While the five constructs of the model have been widely utilized and validated, the items applied to measure them need to be adapted to the context of purchasing health insurance online with AI. Table 1 illustrates the five constructs and the sources of the items that were adapted. Each question has a seven-point Likert for the participant to give their feedback from strongly disagree (1) to strongly agree (7).

Table 1. Constructs and their indicators.

| Role | Construct | Item | Source of construct items |
|---|---|---|---|
| Enablers | Perceived Ease of Use | PEOU1B1/ PEOU1B2 | (Davis 1989) |
| | Perceived Usefulness | PU1B1/ PU1B2 | (Davis 1989) |
| Barriers | Trust in AI | T1B1/T1B2 | (McKnight, Choudhury, and Kacmar 2002; McKnight et al. 2011) |
| | Personal Information Privacy Concerns from AI | PIPC1B1/PIPC1B2 | (Dinev et al. 2013; Xu et al. 2008) |
| Outcome | Usage of insurance utilizing AI | UI1B1/UI1B2 | (Zhao, Ni, and Zhou 2018; Wu, Li, and Fu 2011; Hoque and Sorwar 2017) |

### Data collection

This research evaluates whether a consumer purchasing health insurance without visible AI will have higher trust and lower PIPC compared to when there is visible AI in the interaction. As this is evaluated from the consumer's perspective, two consumer journeys are used with two separate groups of participants. The two journeys used are based on the process a consumer goes through in five insurers that were evaluated. The first group, ß$_1$, are given a consumer journey for purchasing insurance in a process where AI is not visible and there is human participation from the insurance company. The second group ß$_2$ are given a consumer journey for purchasing health insurance in a process where AI is used at every step and this role of AI is made clear to the consumer. The participants then completed a survey that covers the seven hypotheses. The survey was disseminated using the SoSci Survey platform that meets GDPR requirements and stores the data within the EU. The minimum sample size for a statistical power of 80% and a significance level of 1% for the model of five latent variables with three indicators each was calculated to be 176 (Hair et al. 2014). For the first group 248 participants completed the first survey and for the second group 237 completed the second survey. After incomplete and unreliable surveys were taken out 221 were left for the first survey and 217 for the second. The participants were UK residents. The demographic information for both groups is presented in Table 2.

### Data analysis technique

The model was tested with Structural Equation Modeling (SEM) with the Partial Least Squares method (PLS). The relationships between the five variables was tested for each of the two groups separately and then the two groups were compared between themselves across the seven hypotheses. The two groups were compared by applying PLS multigroup analysis (MGA) and bootstrapping with Smart PLS. First the measurement model is evaluated, followed by the structural model.

Table 2. Demographic information for both groups.

|  | AI not visible (ß$_1$) | AI visible (ß$_2$) |
| --- | --- | --- |
| Gender |  |  |
|   Female | 98 | 106 |
|   Male | 123 | 111 |
| Age |  |  |
|   Under 18 | 11 | 14 |
|   18–24 | 96 | 92 |
|   25–39 | 80 | 74 |
|   40–59 | 26 | 18 |
|   60 or older | 8 | 19 |
| Education level |  |  |
|   Without educational level | 8 | 12 |
|   High school | 77 | 81 |
|   Undergraduate university degree | 80 | 94 |
|   Post-graduate university degree | 56 | 30 |
| Income (in British Pounds) |  |  |
|   No regular income | 13 | 15 |
|   400–1200 | 26 | 16 |
|   1201–3000 | 71 | 85 |
|   3001–5000 | 84 | 81 |
|   > 5000 | 27 | 20 |
| Total | 221 | 217 |

## Data analysis and results

The purpose of the multigroup analysis was to evaluate the effect of the moderator variable, in this case having visible human involvement, between the two groups. The Smart PLS-MGA will show whether the difference is statistically significant. The null hypothesis H0 is that there is no difference between the two groups and the alternative hypothesis H1 is that there is a difference. As PLS-MGA is a new and evolving analysis method, simpler descriptive statistics were also implemented.

### Descriptive statistics

The differences between the mean values of the two groups are not large as illustrated in Table 3. However, two points about the difference between the scenarios must be highlighted: Firstly, they are consistent across the six items forming the two constructs of trust and PIPC. Secondly, in the first scenario, trust is marginally above 4, so marginally positive, and PIPC is marginally below 4. This positive trust and negative PIPC is conducive to a purchase. In the second scenario, despite the small difference this is reversed and therefore trust and PIPC are not conducive toward a purchase. This illustrates how small changes in perceptions on trust and PIPC can nevertheless be decisive.

Table 3. Mean values for both groups.

|  | Perceived Ease of Use | Perceived Usefulness | Trust | Perceived Info. Privacy Concern | Use of Health Insurance |
|---|---|---|---|---|---|
| AI not visible ($\beta_1$) | 3.891 | 3.910 | 4.017 | 3.879 | 3.561 |
| AI visible ($\beta_2$) | 3.739 | 4.000 | 3.704 | 4.158 | 3.670 |

### Measurement model

The reflective measurement model was evaluated in several ways. Table 4 shows the results of the measurement model analysis. The factor loadings are over 0.7 so the indicators appear to be sufficiently reliable. The Composite Reliability (CR) is above 0.7 so the construct reliability between the items and the latent variable is sufficient (Hair et al. 2014). The convergent validity is evaluated by the Average Variance Extracted (AVE) AVE is above the required minimum of 0.5. The discriminant validity is below the 0.85 threshold (Hair et al. 2014).

The tests for measurement invariance are illustrated in Table 5. There is some invariance for two of the fifteen indicators, PEOU-3 and UI-1. Each variable has three reflective items which reduce the influence of each item with some invariance. The influence of a small degree of measurement invariance of an item in PLS-MGA is a topic of debate (Sarstedt, Henseler, and Ringle 2011) with some considered acceptable in multigroup analysis, as different groups may have some difference in their understanding of the measurement model (Rigdon, Ringle, and Sarstedt 2010).

### Structural model

The coefficient of determination $R^2$ for endogenous latent variables is 'weak' for P (0.004, 0.018) and T (0.004, 0.026) 'moderate' for PU (0.339, 0.419) and UI (0.562, 0.633) (Chin 1998). This was the same across both groups. The path coefficients are presented in Table 6. The final column evaluates the difference between the two models and whether the hypotheses are supported. Values below 0.05 or above 0.95 are significant in the PLS-MGA analysis. The paths T-UI (H1) and PU-UI (H6) have a significant difference. There was a difference expected for H1 which is supported. The difference for H6 was not expected. There was also a difference expected for PIPC-UI but none was found. The remaining hypotheses had no differences as expected. Therefore, the hypotheses H1, H3, H4, H5, H7 are supported by PLS-MGA and the hypotheses H2, H6 are not supported.

Table 4. Results of the measurement model analysis.

| Items | | Loadings | CR | AVE | AI not visible/AI visible | | | | |
|---|---|---|---|---|---|---|---|---|---|
| | | | | | | | Discriminant validity heterogeneity | | |
| | | | | | PU | PEOU | U | T | PIPC |
| Perceived Usefulness | PU-1 | 0.836/0.818 | 0.899/0.882 | 0.748/0.714 | 0.720/0.833 | | | | |
| | PU-2 | 0.877/0.827 | | | | | | | |
| | PU-3 | 0.881/0.889 | | | | | | | |
| Perceived Ease of Use | PEOU-1 | 0.782/0.735 | 0.862/0.838 | 0.676/0.633 | | | | | |
| | PEOU-2 | 0.803/0.857 | | | | | | | |
| | PEOU-3 | 0.879/0.792 | | | | | | | |
| Use of Health Insurance | UI-1 | 0.861/0.773 | 0.874/0.869 | 0.698/0.690 | 0.416/0.759 | 0.215/0.189 | | | |
| | UI-2 | 0.804/0.851 | | | | | | | |
| | UI-3 | 0.840/0.865 | | | | | | | |
| Trust | T-1 | 0.842/0.891 | 0.896/0.898 | 0.741/0.747 | 0.243/0.402 | 0.084/0.189 | 0.842/0.819 | | |
| | T-2 | 0.840/0.827 | | | | | | | |
| | T-3 | 0.899/0.873 | | | | | | | |
| Perceived Info. Privacy Concern | PIPC-1 | 0.879/0.896 | 0.914/0.923 | 0.779/0.798 | 0.004/0.184 | 0.064/0.189 | 0.592/0.587 | 0.576/0.573 | |
| | PIPC-2 | 0.881/0.916 | | | | | | | |
| | PIPC-3 | 0.888/0.868 | | | | | | | |

Table 5. Test for measurement invariance.

| Items | Outer Loadings-diff (AI not visible-AI visible) | p-Value |
|---|---|---|
| Perceived Usefulness | | |
| PU-1 | 0.019 | .294 |
| PU-2 | 0.050 | .064 |
| PU-3 | 0.008 | .639 |
| Perceived Ease of Use | | |
| PEOU-1 | 0.047 | .225 |
| PEOU-2 | 0.053 | .886 |
| PEOU-3 | 0.087 | .010 |
| Use of Health Insurance | | |
| UI-1 | 0.087 | .008 |
| UI-2 | 0.045 | .850 |
| UI-3 | 0.026 | .829 |
| Trust | | |
| T-1 | 0.048 | .969 |
| T-2 | 0.013 | .358 |
| T-3 | 0.025 | .128 |
| Perceived Info. Privacy Concern | | |
| PIPC-1 | 0.016 | .789 |
| PIPC-2 | 0.035 | .909 |
| PIPC-3 | 0.019 | .227 |

Table 6. Multi-group comparison test results.

| Path | Coefficients AI not visible | AI visible | Hypotheses | PLS-MGA: p-value ($\beta_1$ vs $\beta_2$) |
|---|---|---|---|---|
| T-UI | 0.536 | 0.414 | H1: $\beta_1 > \beta_2$ | .048 |
| PIPC-UI | −0.221 | −0.205 | H2: $\beta_1 < \beta_2$ | .576 |
| PEOU-T | 0.066 | 0.163 | H3: $\beta_1 = \beta_2$ | .827 |
| PEOU-PIPC | 0.060 | 0.135 | H4: $\beta_1 = \beta_2$ | .762 |
| PEOU-PU | 0.582 | 0.647 | H5: $\beta_1 = \beta_2$ | .858 |
| PU-UI | 0.237 | 0.443 | H6: $\beta_1 = \beta_2$ | .994 |
| PEOU-UI | 0.010 | −0.019 | H7: $\beta_1 = \beta_2$ | .372 |

## Discussion

### Theoretical contributions

AI is having an increasing influence and the health insurance sector is both intrinsically important and likely to be an important benchmark for the satisfactory exchange of sensitive information with AI. This research brought together the literature of AI, insurance, health, e-commerce, trust and Personal Information Privacy Concerns (PIPC). It is important to understand the consumer's perspective as they have beliefs on what the role AI should be. Furthermore, the insurer must explicitly inform them if they are using AI when communicating with them in the interface (European-Commission 2019). This research identified two scenarios, one with limited AI that is not in the interface, whose presence is not explicitly revealed to the consumer and a second scenario where there is an AI interface and AI evaluation, and this is explicitly revealed to the consumer.

The two scenarios were modeled and compared using SEM PLS-MGA. Both models were similar in terms of which paths were strong and which were weak. The pathways from PEOU to PU, PU to UI, T to UI and PIPC to UI were strong while the paths from PEOU to UI, T and PIPC were weak. This indicates that the model captures most relationships well apart from PEOU which only has a strong influence on PU.

The analysis gives further support to the hypotheses H1, H3, H4, H5, H7. However, hypotheses H2, H6 are not supported. Therefore, both descriptive and PLS-MGA, support the different level of trust with and without visible AI involvement. Furthermore, it is also supported that trust is higher without visible AI involvement. This extends literature on how trust is lower in certain contexts (Thatcher et al. 2011) to the context of health insurance with AI. Our findings also extend the literature on how different forms of AI interaction influence trust (Torre, Goslin, and White 2020). Most of the current literature is focused on the use of language by clearly visible AI so it is beneficial that this research also evaluated the less visible AI. While the mean of PIPC was lower without visible AI involvement, PLS-MGA did not identify this as statistically significant within the model. Therefore, there is some support that privacy concerns that were validated to be stronger in other contexts (Dinev et al. 2015) are also stronger in this context with AI (Park and Shin 2020). More specifically, there is some support that privacy concerns influence the sharing of personal information on health (Park and Shin 2020; Chen, Zarifis, and Kroenung 2017).

## Implications for practice

The implications for practice are related to how the reduced trust and increased privacy concern with visible AI are mitigated.

### Avoid being explicit about the use of AI

In some parts of the world the use of AI must be stated and visible so there is no choice for health insurers. If there is no legal requirement to be explicit about the use of AI, then the insurer may decide not to be explicit about its use. This strategy has the limitation that several uses of AI such as chatbots and natural language processing are hard to conceal.

### Mitigate the lower trust with explicit AI

The first step to mitigating the lower trust caused by the explicit use of AI is to acknowledge this challenge and the second step is to understand it. The quantitative analysis supports the existence of this challenge and the literature review indicates what causes it. The causes are the reduced transparency and explainability. A statement at the start of the consumer journey about the role AI will play and how it works may reinforce transparency and help to explain it, which will then reinforce trust. Secondly, the level of the importance of trust is increased as the perceived risk is increased. Therefore, the risks should be reduced. Thirdly, it should be illustrated that the increased use of AI does not reduce the inherent humanness. For example, it can be shown how humans train AI and how AI adopts human values. Alternatively, as it has been proven that trust building can focus on human (benevolence, integrity and ability) or system characteristics (helpfulness reliability and functionality) (Lankton, McKnight, and Tripp 2015), trust building can focus on system characteristics.

Lastly, beyond the specific consumer and their journey, society can be influenced to build trust toward AI. In addition to the psychological responses to the specific consumer journey, there are also the social influences. Therefore, society in general can be offered a positive narrative on the trustworthiness of AI and share positive experiences in social learning.

### Mitigate the higher personal information privacy concern (PIPC) with explicit AI

The consumer is concerned about how AI will utilize their financial, health and other personal information. Health insurance providers offer privacy assurances and privacy seals, but these do not explicitly refer to the role of AI. Assurances can be provided about how AI will use, share and securely store the information. These assurances can include some explanation of the role of AI and cover confidentiality, secrecy and anonymity. For example, while the consumer's information may be used to train machine learning it can be made clear that it will be anonymized first. The consumer's perceived privacy risks can be mitigated by making the regulation that protects them clear.

### Limitations and future research direction

Many theories and models of trust have been found to be valid across different cultures, however, it has also been proven that there can be some variation across cultures (Connolly 2013). Therefore, the model developed here could be further explored in different cultures.

While it was shown that trust is different with and without visible AI the role of PIPC should be explored further. The differences identified offer new avenues for further exploration. Furthermore, the value and limitations of SEM PLS-MGA were visible in this type of methodology.

### Conclusion

This research identified two consumer journeys for purchasing health insurance: The first has limited AI that is not visible in the interface and its presence is not explicitly revealed to the consumer. The second has AI in the interface and the evaluation, and its presence is explicitly revealed to the consumer.

The two scenarios were modeled and compared using SEM PLS-MGA. For both models Perceived Usefulness, Trust and Personal Information Privacy Concern (PIPC) influenced the use of health insurance. Both descriptive analysis and PLS-MGA, support the lower level of trust with visible AI involvement in comparison to when AI is not visible. The mean of PIPC was higher with visible AI but this was not statistically significant within the model. These contributions clarify the relationship between the consumer, AI and the health insurance provider and set an agenda for future research on this topic. This agenda might be extended beyond health insurance to other transactions and applications, particularly those that require sensitive information.

## ORCID

Alex Zarifis  http://orcid.org/0000-0003-3103-4601
Peter Kawalek  http://orcid.org/0000-0002-6248-7745
Aida Azadegan  http://orcid.org/0000-0002-6069-8610

82 A. ZARIFIS ET AL.